# How to Achieve High Spatial Resolution in Organic Optobioelectronic Devices?

Luca Fabbri[1], Ludovico Migliaccio[2], Aleksandra Širvinskytė[1], Giacomo Rizzi[1], Luca Bondi[1], Cristiano Tamarozzi[1], Stefan A.L. Weber[3], Beatrice Fraboni[1], Eric Daniel Glowacki[2], Tobias Cramer*[1]

[1]Department of Physics and Astronomy, University of Bologna, Viale Berti Pichat 6/2, 40127 Bologna, Italy

[2]Bioelectronics Materials and Devices Lab, Central European Institute of Technology, Brno University of Technology, Purkyňova 123, 61200, Brno, Czech Republic

[3]Institute for Photovoltaics, University of Stuttgart, Pfaffenwaldring 47, 70569 Stuttgart, Germany

**Abstract**

Light activated local stimulation and sensing of biological cells offers enormous potential for minimally invasive bioelectronic interfaces. Organic semiconductors are a promising material class to achieve this kind of transduction due to their optoelectronic properties and biocompatibility. Here we investigate which material properties are necessary to keep the optical excitation localized. This is critical to single cell transduction with high spatial resolution. As a model system we use organic photocapacitors for cell stimulation made of the small molecule semiconductors $H_2Pc$ and PTCDI. We investigate the spatial broadening of the localized optical excitation with photovoltage microscopy measurements. Our experimental data combined with modelling show that resolution losses due to the broadening of the excitation are directly related to the effective diffusion length of charge carriers generated at the heterojunction. With additional transient photovoltage measurements we find that the $H_2Pc$/PTCDI heterojunction offers a small diffusion length of $\lambda_d = 1.5 \pm 0.1$ µm due to the small mobility of charge carriers along the heterojunction. Instead covering the heterojunction with a layer of PEDOT:PSS improves the photocapacitor performance but increases the carrier diffusion length to $\lambda_d = 7.0 \pm 0.3$ µm due to longer lifetime and higher carrier mobility. Furthermore, we introduce electrochemical photocurrent microscopy experiments to demonstrate micrometric resolution with the pn-junction under realistic aqueous operation conditions. This work offers valuable insights into the physical mechanisms governing the excitation and transduction profile and provide design principles for future organic semiconductor junctions, aiming to achieve high efficiency and high spatial resolution.

**Introduction**

Organic semiconductors are attracting much interest as photoactivated biological transducer materials in recent years.[1,2] The interest is sparked by a range of different properties all combined in this class of materials concerning optoelectronic, electrochemical and biological functionalities.[3,4] The central objective of this research is to develop implantable thin film devices that transduce an optical stimulation into a physico-chemical response, that is able to impact cellular behavior. In contrast to electrically operated biomedical devices, such organic optobioelectronic devices offer the advantage of wireless operation and are therefore less invasive when implanted in the body. Photons with a wavelength in the tissue transparency window can then activate the device allowing full control of the transduction in time and in space. Possible biomedical applications regard retinal implants to restore vision,[1,5] wireless peripheral nerve interfaces,[6] optically activated scaffolds for guided cell differentiation or smart in-vitro systems enabling optically controlled stimulation or silencing of excitable cells.[7,8] Examples in literature aiming at these applications have tested different organic semiconductors[9] such as single layers of semiconducting polymers, semiconducting polymer based bulk heterojunctions or small molecule semiconductor planar heterojunctions.[10,11] The choice of the semiconductor materials and the device architecture yield different physico-chemical mechanisms for the phototransduction process. Depending on the fate of the excitons and separated charge carriers generated in the semiconductor, one distinguishes photocapacitive, photoelectrochemical or thermal transduction mechanisms. [9,12]

A central advantage of optobioelectronic devices is the possibility to control the area of transduction by the shape of the light pulse. A focused light beam can potentially transduce signals to a single biological cell without requiring any microstructuring of the semiconducting thin film into arrays of pixels.[13] Such an advantage would significantly reduce the complexity of devices such as artificial retinas. In addition, activated areas can adapt in real-time to follow the shape of the dynamic biological system interfaced to the device. To achieve this vision, high requirements on the spatial resolution of the optoelectronic transduction process have to be fulfilled. Basic reasoning clarifies that the spatial resolution in optobioelectronic interfaces is defined at three different levels. First, the size of the light spot, that activates the semiconductor matters. Theoretically, here only limitations due to the diffraction limit occur and sub-cellular resolution can be achieved. Second, excitons and separated charge carriers generated in the semiconducting layer are subjected to diffusion and hence cause a broadening of the excitation that reduces the spatial resolution. In crystalline inorganic semiconductors the diffusion length extends over several tens of micrometers and spatial resolution can only be obtained by structuring the semiconductor into pixels to limit the dispersion of the local excitation.[14–16] Third, once the optical excitation is transduced into a physico-chemical stimulus (i.e. an electric field or a released chemical) further broadening occurs in the biological medium before the signal reaches its target. The smallest resolution of the transduction process is determined by all three contributions. Materials research can optimize organic semiconductor properties to minimize resolution losses. To guide this optimization and to reach the intrinsic resolution limit of organic optobioelectronic interfaces, a detailed understanding of the processes that impact on resolution in organic optobioelectronic devices are necessary.

In this work we investigate the spatial resolution limits in organic optobioelectronic devices with a pn-heterojunction fabricated with the organic semiconductors $H_2Pc$ (Phthalocyanine) and PTCDI (3,4,9,10-Perylenetetracarboxylic diimide). Such photo-capacitor devices have been developed as stable photo-transducers in in-vitro and in-vivo experiments.[11,13] Upon illumination, charge separation generates strong enough electric fields to depolarize neuronal cells in close vicinity to the illuminated spot.[5,17] To investigate the spatial and temporal properties of the excitation profile generated in the planar heterojunction we exploit Kelvin Probe Force Microscopy and transient photovoltage measurements. The findings are analyzed with a carrier diffusion and recombination model. Our results show how carrier mobility and lifetime control the broadening of the excitation in the semiconducting layer. We apply our findings to demonstrate spatial resolution in the range of cellular length scales in devices with unpatterned pn-heterojunction layers in photocurrent mapping experiments. From our findings we deduce design principles for future organic semiconductor junctions combining high efficiency and high spatial resolution.

**Materials and Methods**

Sample Fabrication: Photocapacitor devices were fabricated through physical vapor deposition onto glass/ITO slides. The glass surfaces were subjected to a 10-minute sonication in acetone followed by another 10-minute sonication in isopropanol. The treated glasses were then exposed to oxygen plasma for 5 minutes at 3 sccm and maximum power (Diener plasma cleaner). To enhance the adhesion of subsequent layers, we performed a 2-hour vapor-phase octyltriethoxysilane (OTS) silanization on the ITO layer at 90 degrees Celsius. Following OTS, the substrates were rinsed by sonication in acetone for 10 min to remove any silanization multilayer. The OTS layer was found to improve the adhesion of the organic semiconductor layer and prevent delamination during operation in water, producing reliably higher photovoltage than with bare ITO substrate. The organic semiconductors $H_2PC$ (p-type) and PTCDI (n-type) were evaporated respectively at a rate of 0.5 Å $s^{-1}$ and 5-6 Å $s^{-1}$ at a base pressure of <$10^6$ mbar. The total thickness of the photoactive layers is 60 nm given by 30 nm of $H_2PC$ and 30 nm of PTCDI. The additional PEDOT:PSS layer entails a thickness of approximately 200 nm and it was prepared from commercial PEDOT:PSS PH 1000 mixed with (3-Glycidyloxypropyl) Trimethoxysilane (GOPS) 2% V/V. The solution was first sonicated for 15 minutes and then deposited by spin coating (200 rpm 20 s, 650 rpm 30 s, 2000 rpm 30 s). Further, the sample was baked at 140 degrees Celsius for 30 min and soaked overnight in DI water to remove the small molecular weight components embedded in the polymer.

Electrostatic Force Microscopy: We conducted Kelvin Probe Force Microscopy (KPFM) measurements using the Park NX-10 Atomic Force Microscope (AFM). For this experiment, we employed an NSC36 cantilever (MikroMasch) made of n-type silicon and coated with a layer of chromium and gold. The cantilever has a typical force constant of 2 N/m and resonates at a frequency of ∼70 kHz. The light source for KPFM experiments is a Thorlabs L520P50 laser diode with wavelength $\lambda$=520 nm, operated with a dedicated driver. The optical setup to focus the laser source is home-made with Thorlabs components. The laser was modulated at a frequency of 720 Hz to measure the electrostatic force modulation signal with a Zurich MFLI lock-in amplifier. The Suppl. Mat. describes how this signal is used to obtain the local photovoltage.

Transient Photovoltage Measurements: The sample was placed inside a custom sample holder to allow the presence of a second ITO contact separated from the pn-junction with a thin dielectric. We employed the Thorlabs DC220 LED driver to operate a Thorlabs green LED M505L4 with wavelength $\lambda$=530 nm to uniformly illuminate the sample. The modulation duty cycle was chosen to obtain a pulse time of 50 μs. The transient photovoltage was measured with the oscilloscope integrated in the Zurich MFLI lock-in amplifier.

Modulated electrochemical photocurrent microscopy: The electrochemical photocurrent microscope consists of a two-dimensional actuator system built with two PI (Physik Instrumente) L-406 linear stages controlled by two C-663 drivers connected in daisy chain mode. A vertical alignment system is fixed on the uppermost stage and allows for a precise focusing of the laser beam on the photoelectrochemical (PEC) cell. The light source is a Thorlabs L520P50 laser diode with wavelength $\lambda$=520 nm, powered by the Thorlabs

DC220 LED and modulated at 4.4 kHz. The PEC cell is composed of a $H_2Pc$/PTCDI pn-junction on top of which the desired objects can be deposited. Everything is enclosed in a Faraday cage to isolate the system from external noise and allow a correct measurement of small currents (<nA). On top of the pn-junction we placed a PDMS waterproof pool to contain the electrolyte (Phosphate-buffered saline, PBS) which is referenced though an Ag/AgCl electrode. The signal from the pn-junction is amplified with a Femto current amplifier (DHPCA-100) and demodulated with the Zurich MFLI lock-in amplifier.

**Results**

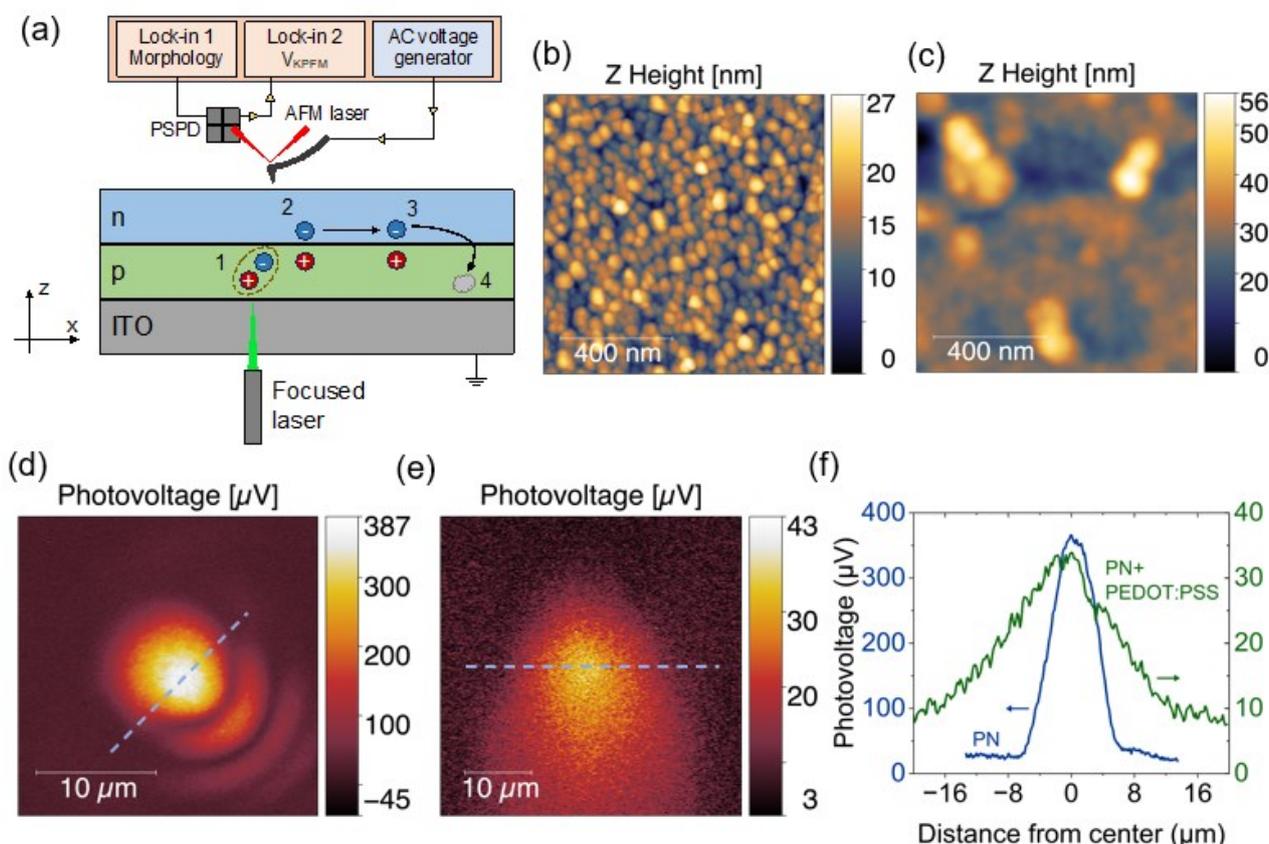

**Figure 1: (a) Schematic of the experimental setup and the physical mechanism causing resolution losses in the semiconductor. (b,c) AFM morphology images of the investigated samples consisting of the organic pn heterojunction without (b) and with (c) a top layer of PEDOT:PSS. (d,e) Photovoltage image of the sample without (d) and with (e) the layer of PEDOT:PSS. (f) Comparison of photovoltage profiles of the two samples.**

Figure 1 introduces to the KPFM experiments performed on the planar $H_2Pc$/PTCDI heterojunction to measure the broadening of the local optical excitation. Illumination is done through the transparent substrate with a laser diode with a wavelength of 520 nm at which both p and n semiconducting layers exhibit strong absorption.[18,19] The laser is focused through a 2 µm pinhole and optical lenses to yield a spot on the semiconductor with a gaussian profile of $\sigma_{SPOT} = (2.59 \pm 0.03)\,\mu m$ (see figure S3 in Suppl. Mat). With KPFM, we measure the surface morphology and the local surface photovoltage defined as $\phi(x) = V_{CPD,light} - V_{CPD,dark}$ where $V_{CPD}$ denotes the contact potential measured at the exposed surface of the heterojunction in darkness ($V_{CPD,dark}$) and during illumination ($V_{CPD,light}$). To improve the signal to noise ratio in the photovoltage measurement, we modulate the light source with a frequency of 720 Hz and measure directly the modulation of the off-resonance electrostatic force microscopy signal (see methods section for details). Typical KPFM images of the investigated samples are shown in figures 1b and 1c. We compare the pure pn-junction to a junction covered with a 200 nm thin film of PEDOT:PSS (poly[3,4-ethylenedioxythiophene]:poly[styrene sulfonate]). Such PEDOT:PSS films have been demonstrated to increase photovoltage and photocurrent generation in heterojunctions.[20–22] The sample without PEDOT:PSS exhibits a morphology with evenly distributed nanometric structures corresponding to nanocrystals formed during the thermal deposition process.[23] Instead the PEDOT:PSS layer displays a different morphology where the roughness of the underlying nanocrystalline film disappears and instead some larger accumulations of polymeric material emerge. The corresponding photovoltage maps are shown in figures 1d and 1e. In both images, the center of the light spot is clearly visible and corresponds to the region with the maximum intensity. In the pure H$_2$Pc/PTCDI sample, the photovoltage remains localized and in some parts the diffraction pattern due to the circular aperture emerges. Instead on the PEDOT:PSS covered surface, the photovoltage signal is more delocalized. This impression is confirmed in figure 1f that provides profiles of the photovoltage maps as indicated (blue dashed lines). In addition to the broadening also a strong reduction in maximum photovoltage value is observed in the presence of PEDOT:PSS.

To analyze our findings, we consider the basic physical processes determining the spatial resolution for optoelectronic excitations in organic p-n heterojunctions as depicted in figure 1a. Excitons are generated in the semiconducting layer in the focused light spot (1) and dissociate at the p-n interface into charge separated states (2). As positive and negative charges accumulate in the respective p and n layer, a shift in the vacuum level happens and the photovoltage is built-up. Photogenerated charges are subjected to diffusive motion along the heterojunction and charge spreads out of the illuminated area (3) causing a reduction in spatial confinement of the optoelectronic excitation. Finally, recombination (4) reduces the local concentration of photogenerated carriers. Combining these processes, the local variation of charge carrier density $n(\vec{x},t)$ is given by

$$\frac{\partial n(\vec{x},t)}{\partial t} = G(\vec{x},t) + D\nabla^2 n(\vec{x},t) - \frac{n(\vec{x},t)}{\tau(n)} \tag{1}$$

where $G(\vec{x},t)$ is the generation rate due to the incident light, D is an effective charge carrier diffusion constant and $\tau(n)$ is the lifetime of charge carriers whose dependence on carrier concentration is governed by the type of recombination mechanism. As usual we introduce the carrier diffusion length as $\lambda_d = \sqrt{D\,\tau(n)}$. In general, the charge carrier concentration $n(\vec{x},t)$ is a function of the three spatial coordinates. For the case of a planar heterojunction made with organic semiconductors with low doping concentrations, the motion along the z-direction is faster than the planar one. This allows to consider the two components separately and study the x-y diffusion once the charges already moved away from the junction. Further, we neglect drift motion caused by electric fields because at the planar heterojunction the field is perpendicular to the interface. Only at the border of the illuminated spot, stray fields with components in the direction of the junction cause slightly accelerated drift, but it is expected to be small compared to the impact of thermal diffusive motion in disordered organic semiconductors. Equation (1) contains all the fundamental contributions determining the intrinsic spatial resolution limit for an optically excited semiconductor transducer device. In such a device, locally generated charge carriers initiate the biological sensing or stimulation process. But they can also diffuse out of the illuminated area before triggering the transduction event, reducing in this way the spatial resolution.

In order to achieve a quantitative analysis of the KPFM surface photovoltage maps, we assume that the local photovoltage *ϕ(x)* is directly proportional to the local concentration of photogenerated charge *n(x)* at the heterojunction. This assumption is justified considering that each generated charge is paired to a counterbalancing charge across the p-n interface to form a dipole. Such a dipole is characterized by an average distance *l* between the centers of charge which does not depend on carrier concentration or light intensity. Accordingly, we can simplify by approximating that the dipoles in the p-n junction form a plane capacitor for which the distance between the two plates is *l*. Then the photovoltage can be written as $V_p = \sigma/c = \sum q_i \times l/\varepsilon\varepsilon_0 A = n\,q_0\,l/\varepsilon\varepsilon_0$ demonstrating the proportionality between photovoltage $V_p$ and the carrier concentration *n*.

We can further simplify equation (1) due to the radial symmetry of the system and the isotropic diffusion of the photogenerated charges in the direction parallel to the heterojunction. The only coordinate that needs to be considered is *r* describing the distance from the center of the illumination spot. Further, for the KPFM experiment with continuous illumination, we have to consider the steady state solution of equation (1) with $\partial n(r,t)/\partial t = 0$ so that the carrier concentration *n(r)* is only a function of the radial coordinate *r*. In our experiment we are mostly interested in understanding the diffusion processes outside of the illuminated region causing the reduction in resolution. In this region no charge carrier generation occurs and we can set *G=0*. The resulting differential equation that determines the tail of photovoltage profile then reads

$$\frac{d^2V_p(r)}{dr^2} + \frac{1}{r}\frac{dV_p(r)}{dr} - \frac{1}{D\,\tau(V_p)}V_p(r) = 0 \qquad (2)$$

We substitute the charge carrier lifetime with the expression $\tau(V_p) = a^{-2}K_{bi}^{-1}V_p^{-1}$ based on a bimolecular recombination mechanism as determined with transient photovoltage measurements (see below). Equation (2) has no simple analytical solution, and we use a numerical Runge-Kutta scheme to fit the equation to the experimental data. Boundary conditions are given by the experimental photovoltage and its first derivative at the border of the light spot $(r_{spot})$ i.e. $V_p(r_{spot}) = V_0$ and $V_p'(r_{spot}) = V_0'$ as obtained from the experimental data.

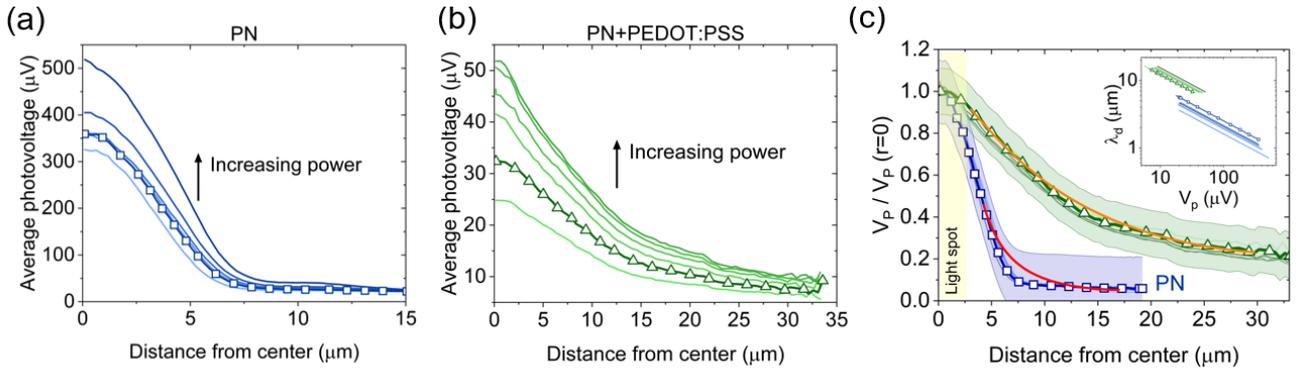

**Figure 2: (a) Averaged photovoltage as a function of distance *r* from the light spot center as measured on the pn junction. (b) same as (a) but for a sample with PEDOT:PSS deposited on top of the pn junction. In both (a) and (b) the photovoltage is reported for different light powers. (c) Average normalized photovoltage profiles of PN and PN+PEDOT:PSS samples with numerical fit to diffusion equation. The inset shows the diffusion length as a function of the local photovoltage $V_P$.**

To compare the model to our KPFM data, we determine averaged radial photovoltage profiles from the images taken at different light intensities. They are shown in figures 2a and 2b for the pure p-n heterojunction (blue squares) and for the one with the top PEDOT:PSS layer (green triangles). Profiles were obtained with different laser powers ranging from 72 nW to 334 nW. In both devices, the photovoltage increases with increasing laser power while the shape of the curves is not affected. In Figure 2c, we present the normalized KPFM photovoltage profiles for both the p-n and p-n-PEDOT:PSS samples as obtained at a power of 126 nW, along with their respective numerical fits using equation (2). The fitting range is determined by the size of the light spot, exclusively considering the region where carrier generation is not present. The boundary conditions are directly calculated from experimental data at the left border of the fitting range. The thinner curves correspond to the other profiles reported in figures 2a and 2b, confirming that the average photovoltage shape remains consistent across various power values. From our fitting procedure with the diffusion equation we obtain the combined parameter $p^2 = a^2 K_{bi} D^{-1}$ from which we can compute the diffusion length by $\lambda_d = \sqrt{1/pV_p}$ (details described in the Suppl. Inf.). The resulting values are shown in the inset of figure 2c. Values of diffusion length at the border of the illumination spot are $\lambda_d(\text{PN}) = (1.5 \pm 0.1)$ μm and $\lambda_d(\text{PN} + \text{PEDOT:PSS}) = (7.0 \pm 0.3)$ μm. The full profile takes the shape of a gaussian bell curve with $V_p \sim \exp\left(\frac{-z^2}{2\sigma_e^2}\right)$ with $\sigma_e^2 = \sigma_{h\nu}^2 + \lambda_d^2$ where the $\sigma_{h\nu}$ describes the gaussian profile of the focused light spot. For the standard deviation $\sigma_e$ describing the profile of the excitation we find $\sigma_e(\text{PN}) = (3.11 \pm 0.01)\mu m$ for the pn-junction and $\sigma_e(\text{PN} + \text{PEDOT:PSS}) = (7.78 \pm 0.08)\mu m$ for the PN+PEDOT:PSS sample.

Our findings demonstrate that the carrier diffusion length in the p-n sample is much shorter than in the p-n-PEDOT:PSS sample. Accordingly the PEDOT:PSS layer causes a broader, more diffuse photovoltage signal. The increased diffusion length is also associated to the lower photovoltage magnitude in the center of the light spot of the PEDOT:PSS functionalized sample. Still, the causes for the extended diffusion length observed in the PN+PEDOT:PSS sample cannot be clarified from these data: an increased diffusion length could be attributed to enhanced charge carrier diffusion but also to extended carrier lifetime.

To answer this question, we perform transient photovoltage measurements that provide access to the carrier lifetime. In such experiments the spatial resolution is not relevant, instead higher requests are put on the temporal resolution of the measurement. Accordingly, we use a macroscopic photovoltage measurement setup as introduced in figure 3a. The sample is mounted on a holder that provides an additional front ITO contact separated from the top layer by a thin dielectric. This structure assures a capacitive coupling between the p-n junction and the additional ITO electrode. An LED ($\lambda$=530 nm) produces light pulses of 50 μs duration illuminating uniformly the heterojunction. The photovoltage transients are amplified with a high impedance amplifier and visualized on an oscilloscope screen. In figures 3b and 3c, we present the recorded photovoltage transients for both the PN and PN+PEDOT:PSS samples obtained at different light power densities ranging from 108 mW cm$^{-2}$ to 531 mW cm$^{-2}$, comparable to the values used in the KPFM experiment. The yellow region in the plots indicates the duration of light illumination. In both samples, we observe a consistent increase in the maximum photovoltage values as power density increases, ranging from approximately 4 mV to 10 mV. Notably, in this experiment the PN+PEDOT:PSS sample exhibits a higher photovoltage response than the pure pn-heterojunction sample.

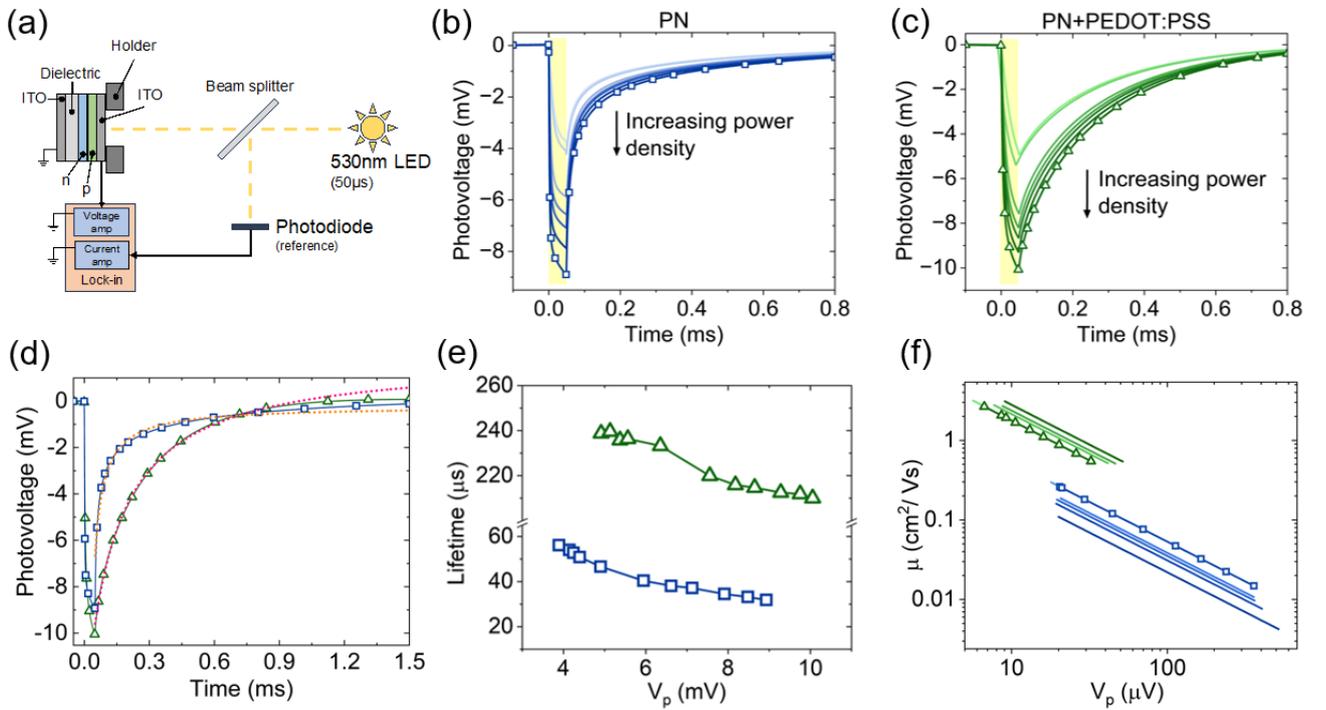

**Figure 3: Photovoltage transient analysis during illumination with 50 μs duration pulses. (a) Schematic of the experimental setup. (b) Photovoltage transients recorded at different power densities for the PN sample and (c) for the pn-junction with the top PEDOT:PSS layer. (d) Transients of both samples obtained at a power density $P$ = 531 mW/cm$^2$ with fit to the model. (e) Average charge carrier lifetime as a function of local photovoltage $V_p$. (f) Mobility of the two samples as a function of laser power (color scale) and local photovoltage.**

In order to analyze the data quantitatively, we consider equation (1) describing generation and recombination processes causing photovoltage transients. In the experiment, we illuminate the whole pn-junction area, therefore the photovoltage is no more a function of position x and we can neglect diffusion processes. Furthermore, we consider only the photovoltage decay when the light source is switched off.

Removing the diffusion and generation terms we obtain the differential equation that describes the photovoltage decay

$$\frac{dn(t)}{dt} = -\frac{n(t)}{\tau(n)} \quad (3)$$

By substituting the characteristic time $\tau$ with the expression for bimolecular recombination and the concentration with photovoltage, we can obtain an analytical solution for equation (3)

$$V_p(t) = \frac{V_{p0}}{1 + V_{p0} b K_{bi} t} \quad (4)$$

where $V_{p0} = V_p(t=0)$ and $b$ is the constant that relates the transient photovoltage to the charge carrier concentration in the p-n junction. In figure 3d, we present the photovoltage transients for both samples measured at a power density of 531 mW cm$^{-2}$, along with their corresponding fits to equation (4). Figure 3e shows the carriers lifetime as function of photovoltage (and thus as function of laser power density) for the two samples. In both cases the lifetime increases when the local photovoltage decreases. Comparing the lifetimes averaged over the photovoltage, we find that in the PN+PEDOT:PSS sample the lifetime is approximately ten times larger than in the p-n sample. With the knowledge of the carrier lifetimes, we can also calculate the local effective diffusivity $D = \lambda_d^2 \tau^{-1}$ from the diffusion length obtained in the KPFM experiment. Using the Einstein relation, we further compute the effective carrier mobility $\mu = D \, k_B^{-1} T^{-1}$. The dependence of mobility on total photovoltage and light intensity is reported in figure 3f where the strong difference between the two samples is clearly appreciable. The average values for the two samples at the border of the light spot are $\mu(\text{PN}) = (0.02 \pm 0.01) \text{ cm}^2 \text{V}^{-1} \text{s}^{-2}$ and $\mu(\text{PN} + \text{PEDOT: PSS}) = (0.58 \pm 0.02) \text{ cm}^2 \text{V}^{-1} \text{s}^{-2}$ showing that PN+PEDOT:PSS sample entails a mobility which is approximately thirty times the one of the pure pn-junction.

Our data allow a better comprehension of the effect of the outer PEDOT:PSS layer on the charge separation and photovoltage generation in photocapacitor devices. We find that PEDOT:PSS increases significantly the lifetime of carriers or in other words it slows down recombination processes. At the same time the effective mobility of generated carriers increases. These effects have different consequences on the operation of photocapacitors: in devices that are fully illuminated, the increased lifetime causes larger photovoltage values and hence improves the performance of the device. In contrast, when devices are only illuminated in a spot, photovoltages are smaller in the presence of PEDOT:PSS as carriers diffuse faster and transfer energy out of the illuminated region. The latter effect would also reduce the resolution of optobiolelectronic interfaces. Knowing the electronic properties of PEDOT:PSS, we try to rationalize why the outer layer of PEDOT:PSS increased the lifetime and effective carrier mobility. Upon illumination negative carriers are accumulated in the outer PTCDI layer of the heterojunction. If these carriers remain close to the interface, they can rapidly recombine with the positive charges in the $H_2Pc$ layer. In the presence of a PEDOT:PSS layer bound to PTCDI layer, negative carriers can be stabilized due to image charges in the highly doped (metallic) PEDOT:PSS layer. Such an image charge effect pulls the generated carriers away from the junction. Additionally, the high doping of PEDOT:PSS enables that some negative charges are transferred into PEDOT:PSS where they fill the mobile hole states. The consequence is that generated carriers are even more distant from the junction and hence less likely to recombine with positive carriers in the $H_2Pc$ layer. Furthermore, space charge present in PEDOT:PSS is expected to have a much higher effective mobility. Values from transport measurements in PEDOT:PSS demonstrate mobilities of up to 10 $cm^2$ $V^{-1}$ $s^{-1}$ [24] exceeding by more than an order of magnitude mobilities associated to PTCDI or $H_2Pc$.

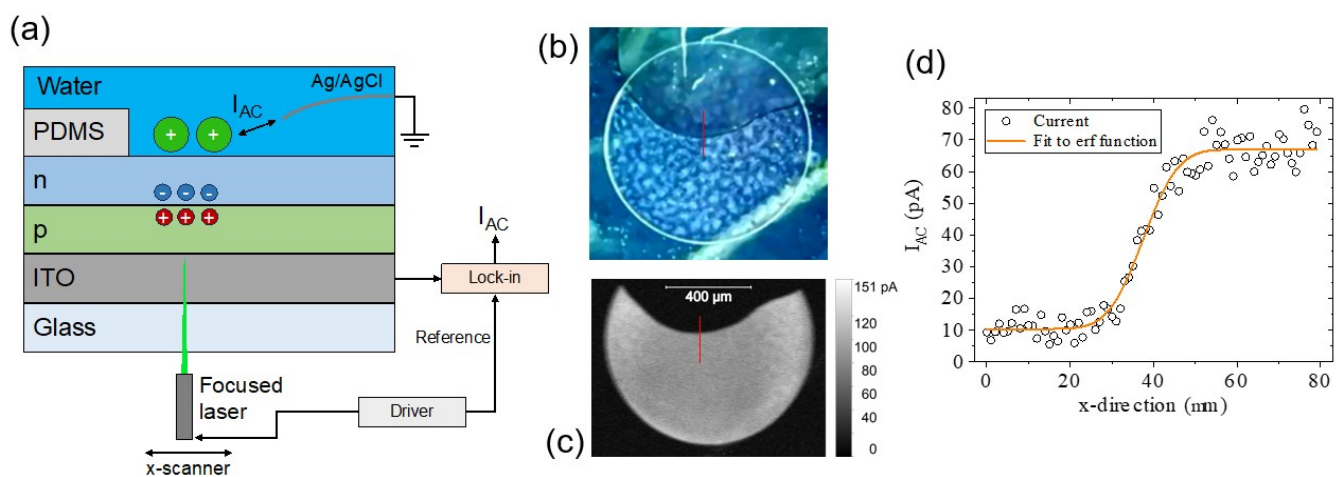

**Figure 4: Photoelectrochemical current microscopy with organic planar p-n junction to determine the spatial resolution. (a)** Scheme of the experiment in which a part of the pn-junction is covered by a dielectric PDMS layer **(b)** Optical micrograph of the p-n-junction electrode covered partially with PDMS. **(c)** AC photoelectrochemical current image of the electrode. **(d)** Profile of the current map and fit to error function. The red lines in (b and c) show the location of the profile.

Our measurements demonstrate that the organic pn-junction is advantageous for interfaces where a high spatial resolution is important. Due to the small diffusion length in the organic heterojunction, the optoelectronic excitation remains localized and a spatial resolution in the order of 3 μm is feasible following the results from the KPFM photovoltage measurements. For bioelectronic applications, photocapacitors are operated in electrolyte and additional mechanisms reducing the spatial resolution could be present. For example, water is a strongly polarizable medium and impacts on optoelectronic processes in the semiconductor.[25] In addition, the electric field diverges in the semi-infinite space of the electrolyte and high resolution requires that transduction processes occur in close vicinity to the semiconductor/electrolyte interface. To address such issues and to demonstrate the potential to achieve high resolution we perform a model imaging experiment with the semiconductor in contact with an aqueous electrolyte. Figure 4a introduces the setup for this photoelectrochemical current imaging experiment. The observed signal is the alternating photocurrent generated by the illumination spot focused on the pn-junction. For the case of a light source with modulated intensity ($f$ = 4.4 kHz) such a photocapacitive alternating current builds up, as the pn-junction is patterned on top of an ITO electrode connected to ground. Upon modulated illumination, the changing polarization of the heterojunction generates an ionic displacement current in the electrolyte entering the Ag/AgCl wire to close the circuit. As the light source is mounted on a x-y stage, we can scan the light spot over the semiconductor sample. Due to the very small illuminated area, the photocurrent signal is small (< nA) and its amplitude and phase are measured with a lock-in amplifier. Similar electrochemical photocurrent microscopy setups have been used to monitor local electrochemical currents at inorganic semiconductor/electrolyte interfaces.[26,27]

The electrochemical photocurrent image of the pn-junction electrode with a diameter of 500 μm is presented in figure 4b and compared to the optical microscopy image. One clearly notices the shape of the circular pn-junction corresponding to the region where photocurrent is generated. A part of the circular electrode is covered by an isolating layer of polydimethylsiloxane (PDMS). Even though the pn-junction is present below the PDMS, no photocurrent is generated in this area. Figure 4d shows a profile of the photocurrent amplitude taken across the border of the PDMS layer along with the corresponding fit to the cumulative error function. The profile shows a clear transition from the covered to the uncovered regions, revealing a gaussian shaped excitation profile described by $\sigma_{e,PDMS} = (6.5 \pm 0.9)\ \mu m$ as extracted from the error function fit. The finding demonstrates that a micrometric resolution can be obtained with the pn-junction in realistic aqueous operation conditions. The resolution value is similar to the size of single cells (~ 10 μm diameter) and shows that single cell transduction is feasible in optically activated organic bioelectronic devices. Compared to the KPFM experiment, the size of the excitation profile doubled in the electrochemical conditions. We associate the reduced resolution at the PDMS/electrolyte interface to the effects of water polarization on carrier lifetime and diffusion as well as possible undercuts or structural inhomogeneities at the PDMS border and related current dispersion. This finding demonstrates that the overall extension of the excitation at such semiconductor/electrolyte interfaces is determined by three contributions $\sigma_e = \sqrt{\sigma_{hv}^2 + \lambda_d^2 + \sigma_{ec}^2}$ in which $\sigma_{hv}$ denotes the extension of the focused light spot, $\lambda_d$ the effective diffusion length in the semiconducting layer and $\sigma_{ec}$ additional losses in the semi-infinite half space of the electrolyte.

**Conclusions**

In this work we provide insight into the physical mechanisms that determine the spatial resolution in organic optobioelectronic devices. As model system we investigate organic photocapacitor devices that consist in a planar organic pn junction made of the small molecule semiconductors H$_2$Pc and PTCDI. Local photovoltage measurements with Kelvin-Probe Force Microscopy show how a focused optoelectronic excitation of a laser spot broadens due to carrier diffusion. By modelling the diffusion and recombination processes we show that the resolution losses due to the broadening are directly related to the effective diffusion length λ$_d$ of carriers in the organic semiconductor. In our experiments we determine a diffusion length of λ$_d$ = (1.5 ± 0.1) μm for the pn-junction and of λ$_d$ = (7.0 ± 0.3) μm for the pn-junction with a top coating of PEDOT:PSS. Further insight into the differences between these two layers is provided with time-resolved photovoltage measurements. The PEDOT:PSS coating increases the carrier lifetime and makes the photocapacitor more effective in terms of charge carrier generation. However, the PEDOT:PSS coating increases also significantly the carrier mobility ($\mu(PN) = (0.02 \pm 0.01)\ cm^2\ V^{-1} s^{-1}$ vs $\mu(PN + PEDOT:PSS) = (0.58 \pm 0.02)\ cm^2\ V^{-2} s^{-1}$) explaining the increased diffusion length and associated losses in resolution.

To demonstrate the strongly localized excitation and the high resolution predicted for the pure H$_2$Pc/PTCDI photocapacitor also under electrochemical conditions, we developed a microscopy experiment in which we map the alternating ionic current generated locally when the pn-junction is immersed in electrolyte and illuminated by a focused laser spot with modulated light intensity. The localization of the photoactivated current source is quantified by measuring how the current is blocked at the border of a dielectric layer covering the heterojunction partially. We determine, under these conditions, a gaussian shaped excitation profile with $\sigma_e = (6.5 \pm 0.9)$ μm. The finding demonstrates that organic small molecule heterojunctions enable to maintain the excitation localized at the scale of single biological cells. With the detailed measurements we can quantify how the overall resolution is determined by the extension of the focused illumination spot, the carrier diffusion length in the semiconductor junction and additional losses due to dispersion of the current in the electrolyte.

**Acknowledgements**

This research was supported by EU Horizon 2020 FETOPEN-2018-2020 Programme "LION-HEARTED", grant agreement n. 828984 (B.F., T.C.). Sample fabrication was supported by CzechNanoLab Research Infrastructure financed by MEYS CR (LM2023051). The authors L.M. and E.D.G. acknowledge funding from the European Research Council (ERC) grant agreement no. 949191.